# Thermoelectric Sensors as Microcalorimeter Load

L. Brunetti[1], L. Oberto[1,2] and E. Vremera[3]

[1]Istituto Nazionale di Ricerca in Metrologia (INRiM)

Strada delle Cacce 91, 10135 Torino, Italia

Tel: + 39 (0)11 3919323, Fax: +39 (0)11 346384, E-mail: brunetti@inrim.it

[2]Politecnico di Torino

Corso Duca degli Abruzzi 24, 10129 Torino, Italia

Tel: + 39 (0)11 3919321, Fax: +39 (0)11 346384, E-mail: oberto@inrim.it

[3]Technical University of Iasi, Faculty of Electrical Engineering

Bld. Dimitrie Mangeron 53 700050, Iasi, Romania

Tel: +40 232 278683; Fax: +40 232 237627; E-mail: evremera@ee.tuiasi.ro

***Abstract***

**Thermoelectric power sensors can now be used as transfer standards, instead of bolometers, in the microcalorimeter technique. This alternative has the technical advantages to be less sensitive to absolute temperature and not downward frequency limited. At INRiM the high frequency power standards are now based on coaxial thermocouples from dc to 34 GHz. Modified commercial thermocouple mounts in 7 mm and 3.5 mm coaxial line are used to realize the national power standard with an accuracy ranging from 0.03 % to 1 % in the mentioned frequency range.**

# I. Introduction

The microcalorimeter technique is always the key technique for the realization of the *primary power standards* at high frequency (HF). The measuring system was originally developed for bolometric sensors based on thermistors or barretters that measure the HF power by means of the *dc power substitution method*. This allows easily tracing HF power standards to the direct current standard, that is a fundamental SI quantity, [1].

Bolometers have some very useful features like that to be free of linearity error and to allow easy measurement of the HF power converted to heat in the bolometer mount, but they have also some drawbacks, anyway. For technical reasons, they are not usable below 10 MHz, are very sensitive to the environmental absolute temperature variations and last, they are now not easily available as commercial devices, especially in waveguide configuration. An alternative power sensor to use with microcalorimeter technique is therefore necessary.

In our knowledge, indirect heating thermocouples of commercial realization have been proposed as transfer standards and microcalorimeter thermal load by PTB for the first time [2]. The device is a true r.m.s. power sensor usable to trace HF power standard to the dc standard without the limitation of the classic bolometers and even more efficiently.

When, at INRiM (formerly IEN), a new microcalorimeter was introduced to operate above 18 GHz, the design was addressed for using thermoelectric power sensors [3] rather then bolometers.

Since then, the system has been improved both in the hardware and in the calibration procedures up to the actual conditions that we are going to present in this paper. Basically, our microcalorimeter operates with coaxial thermocouple power sensors fitted by PC-N and PC-3.5 connectors in a frequency range that spans from dc to 34 GHz or to 40 GHz if K-connector is used.

## II. MEASUREMENT SYSTEM

The INRiM microcalorimeter is a twin-type adiabatic calorimeter, as schematized in Fig.1. The twin-line solution has been used, after having experimented for long time the single inset line, because it is more effective in filtering the thermal disturbances coming from outside even though its realization is more complex. The thermostat is a triple wall container, temperature stabilized by Peltier elements that act only on the intermediate wall, while the other two work like passive thermal shields. This solution allows obtaining temperature stabilization better than 10 mK. The system may operate both with bolometric power sensors and thermoelectric devices but the calibration procedure has been definitively refined for the last type obtaining, with this, a high accuracy improvement of the power standard.

## III. CALIBRATION AND MEASUREMENT THEORY

Microcalorimeter was originally introduced for measuring the bolometer mount *effective efficiency* $h_e$ as defined in [4]. For thermoelectric devices, the formerly definition is no more consistent. It may be conveniently refined as the ratio between a reference (RE) power $P_{RE}$, in dc or low frequency (LF), entering the sensor and the absorbed RF power $P_{RF}$ that produces the same sensor output ($U = const$).

Like it has been adopted in [5]:

$$h_e = \left. \frac{P_{RE}}{P_{HF}} \right|_{U=const.} \tag{1}$$

Formula (1) is only a definition and it cannot be directly used in the measurement process. An utilizable expression for $h_e$ may be determined starting from the fundamental electro-thermal equation of the microcalorimeter, that is an application of the superimposition principle of the linear effects in our measurement system [1], [3]:

$$e = aR\left(K_1 P_{inS} + K_2 P_{inL}\right) \tag{2}$$

In it, $e$ is the thermopile output voltage, $P_{inS}$ the total power dissipated in the sensor mount, $P_{inL}$ the feeding line losses and $K_1$, $K_2$ coefficients describing the power separation between the sensor and feeding line. The dimensional coefficients $a$ and $R$ will disappear in the following because only voltage ratios have to be considered. By supplying the measurement channel of the microcalorimeter with RF power the system response $e_1$ is given by:

$$e_1 = aR\left(K_1 P_{inS} + K_2 P_{inL}\right), \tag{3}$$

while, after the substitution of HF-power with dc/LF power, the thermopile output $e_2$ will be given by:

$$e_2 = aR\left(K_1 P_{inS}\right)_{RE} \tag{4}$$

This expression results because dc/LF losses in the insulating line are assumed and have been experimentally proved to be negligible. If the power substitution maintains the output of the power sensor $U$ to be constant, we can easily obtain an expression containing $h_e$ from the ratio of the thermopile outputs $e_1$ and $e_2$, each one taken at the thermodynamic equilibrium condition. Simplifying and insulating the effective efficiency term, we get:

$$h_e = \frac{e_2}{e_1}\left(1 + \frac{K_2}{K_1}\left(\frac{P_{inL}}{P_{inS}}\right)_{HF}\right) = e_R g \tag{5}$$

So far we have obtained an operative expression of $h_e$ conformal to the definition given in (1).

Equation (5) is usable for the power sensor calibration if the term $g$, dependent on the microcalorimeter characteristics, may be evaluated at each measurement frequency, that is, if the measurement system is calibrated. The authors showed in [6] that the best microcalorimeter calibration process is obtainable if the power sensor, or a twin copy of it, can be reduced to a completely reflective RF-load. In this case, we can set $h_e$ to unity and, reversing (5), $g$ can be determined. However (5) may be still converted in an equivalent formula more convenient from the experimental point of view.

Assuming to perform two identical measurement series, one with the sensor in normal condition and the other with the same in highly reflective condition, we can have enough information to write:

$$h_e = \frac{e_2}{e_1 - e_{1SC}} \tag{6}$$

where $e_{1SC}$ is the thermopile response when a half of the RF power producing $e_1$ is supplied to the power sensor transformed into a reflective load. The RF power reduction of 3 dB is necessary to compensate for the losses produced by the backward traveling wave.

The compact expression (6) is function of the microcalorimeter thermopile output $e$ only, nevertheless it includes all the terms relevant and influent for the power sensor calibration. The term $e_{1SC}$ is clearly a corrective dependent on the feeding line losses.

This discussion and, in some sense, result is valid only if the feeding line losses and the inner temperature of the microcalorimeter are maintained constant, if the power dissipated in the sensor remains steady through a proper power substitution and if it is possible to measure the thermopile response at the thermodynamic equilibrium even when the losses are very small (< 1 nW).

Constance of feeding line losses is achieved with the stabilization of the injected HF power at the system input port. It is done using an analogical leveling loop based on resistive power splitter and a crystal detector. The LF (1 kHz) power that substitutes HF power is, conversely, controlled via a real-time digital system that maintains the output $U$ of the sensor at the same level of the mean value measured in the first HF cycle. This power control method allows the power dissipated into the sensor to remain almost steady as Fig. 2 shows. It can also be seen from the figure that the stabilization achieved on the LF power by means of the digital control system (of the order of 1 μW) is better than the one obtained by the HF leveling loop (some μW).

As can be seen in the pictorial cross section of microcalorimeter in Fig. 3, the temperature of the thermostat is stabilized with Peltier elements driven by a dedicated PID controller and a PT100 thermometer placed on the inner metallic shield. The outer shield isolates the microcalorimeter core from the environment temperature drifts, while the inner shield attenuates the temperature oscillations still produced by PID controller around the setting point.

To check the influence of the residual PID oscillations on the temperature of the measurement chamber, we monitored the microcalorimeter thermopile output without injected signal. Variation of this dc signal can only came from temperature modifications inside the measurement chamber, indeed. Figure 4 shows a scatter plot of the thermopile output versus the PT100 output from which it can be seen that no correlation is present between the two signals. Because there are not other possible influence quantities on the thermopile in this situation, we can assume that the measured variation are only due to the thermal noise of the thermopile itself and that the stabilizing effect of the inner shield is effective. A comparison between the calculated thermal noise of the thermopile and the measured signal confirms this hypothesis furthermore.

To cope with the last requirement previously evidenced, the signal of the thermopile is amplified by a high stability, ultra-low noise amplifier.

It is now important to underline that, for a coherent determination of $e_{1SC}$, it is necessary to change the electromagnetic behavior of the microcalorimeter load without changing the thermal equilibrium. In particular, we need to substitute the sensor with a thermally equivalent short circuit and this needs a particular hardware arrangement.

## IV. HARDWARE ARRANGEMENT

The arrangement that proved to be efficient and easy to use for a correct determination of the corrective term $e_{1SC}$ is shown in Fig. 5.

Two low loss line sections have been inserted between the microcalorimeter test-ports and the power sensors realizing the thermal load. The sections, one of which is a short circuit, are interchangeable between the measurement and reference channel. This allows obtaining the reflectivity condition necessary for calibrating the system without changing significantly the system thermal equilibrium. The short circuit plane is slightly shifted with respect to the original sensor input plane, but this is not so influent because the requirement of lossless short is much more important. Negligible losses are obtained with a precise mechanical machining.

Another hardware improvement concerns the position of the microcalorimeter thermopile. As Fig. 5 shows, it has been moved to the base of power sensor input connector, a position that allows a more efficient detection of the parasitic losses of the thermocouple mount.

Furthermore, as marginal improvement, an additional insulation section has been inserted at the external shield level, obtaining a reduction of the signal noise.

## V. SOFTWARE IMPROVEMENT

Another improvement concerns directly the data analysis. If in the past $e_1$, $e_2$ and $e_{1SC}$ entered (6) were the instrument readings at the thermal equilibrium, now they are more efficiently determined through a statistic process involving the Levenberg-Marquardt algorithm already used in [6]. Basically we record the thermopile output continuously when cycling the HF power with the RE-power, obtaining a saw-tooth signal that we fit with a suitable exponential function whose asymptotes will be the values to enter (6). An example of this waveform is showed in Fig. 6. The last fitting procedure splits the saw-tooth waveform into its basic components: the increasing and decreasing exponentials.

These functions are separately fitted to find their asymptotes, which are, then, averaged before obtaining the asymptotic values to use in (6).

## VI. UNCERTAINTY IMPROVEMENT

Extensive measurements have been performed from 10 MHz to 26.5 GHz in coaxial line with PC-3.5 connector to test the improvements in the measurement set-up.

The result is that we have pulled down the 1 % uncertainty threshold in all the measurement range. In particular, considering the Calibration Factor *K* of the transfer standard, defined as:

$$K = h_e\left(1 - |G|^2\right) \tag{7}$$

where *G* is the reflection coefficient of the standard, we obtain a maximum relative uncertainty ($k$ = 2) of 0.22 % as Fig. 7 shows.

Being conservative, we can claim an expanded relative uncertainty ($k$ = 2) that is includible in the interval (0.03 - 0.5) % for frequencies between 10 MHz and 26.5 GHz. Preliminary measurements show also an upper uncertainty limit of 1 % for frequencies up to 34 GHz and for 3.5 mm coaxial lines.

## VII. CONCLUSIONS

We have presented the last advancements of the microcalorimeter technique at INRiM. A calibration method previously elaborated for bolometers has been extended to thermocouples obtaining a simplified and elegant theory. A refined data analysis technique is extensively applied for realizing the new HF power standards based on thermocouples. The total expanded uncertainty ($k$=2) relative to the Calibration Factor *K* of the HF-power standard can be included in the range (0.03 - 1) % for frequency up to 34 GHz and measurements on 3.5 mm coaxial lines.

## Captions

Fig. 1: Basic structure of the twin type microcalorimeter scheme.

Fig. 2: Power stability in the system; input signal is switched between RF and LF every 90 minutes starting with RF.

Fig. 3: Detailed microcalorimeter cross section.

Fig. 4: Scatter plot of the thermopile output versus the PT100 thermometer output.

Fig. 5. Details of the twin type microcalorimeter load.

Fig. 6: An example of the microcalorimeter thermocouple record together with its fitting curve.

Fig. 7: Relative Uncertainty ($k$ = 2) of the Calibration Factor $K$ of a typical INRiM power transfer standard based on thermocouple.

**Figure 1**

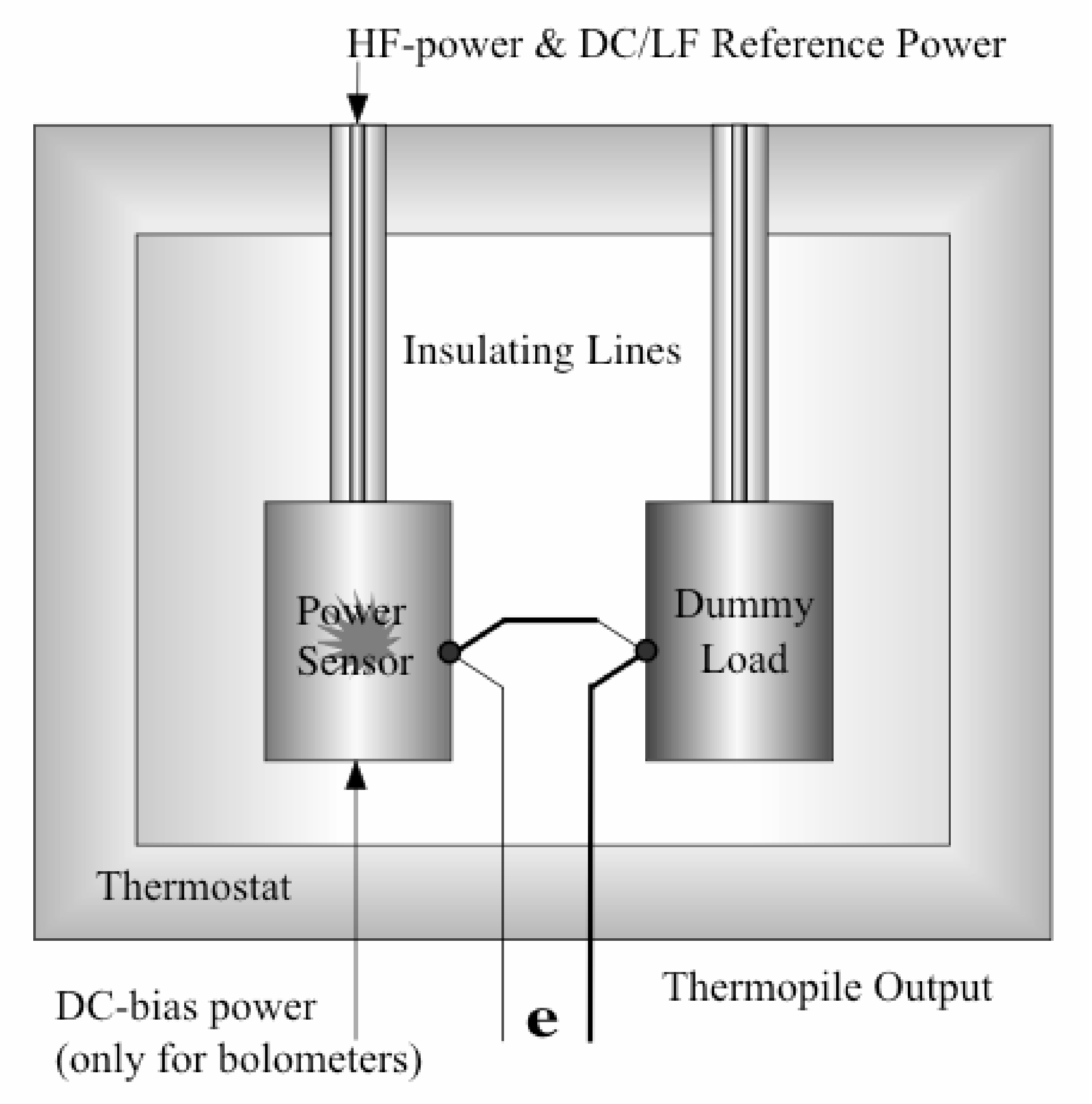

**Figure 2**

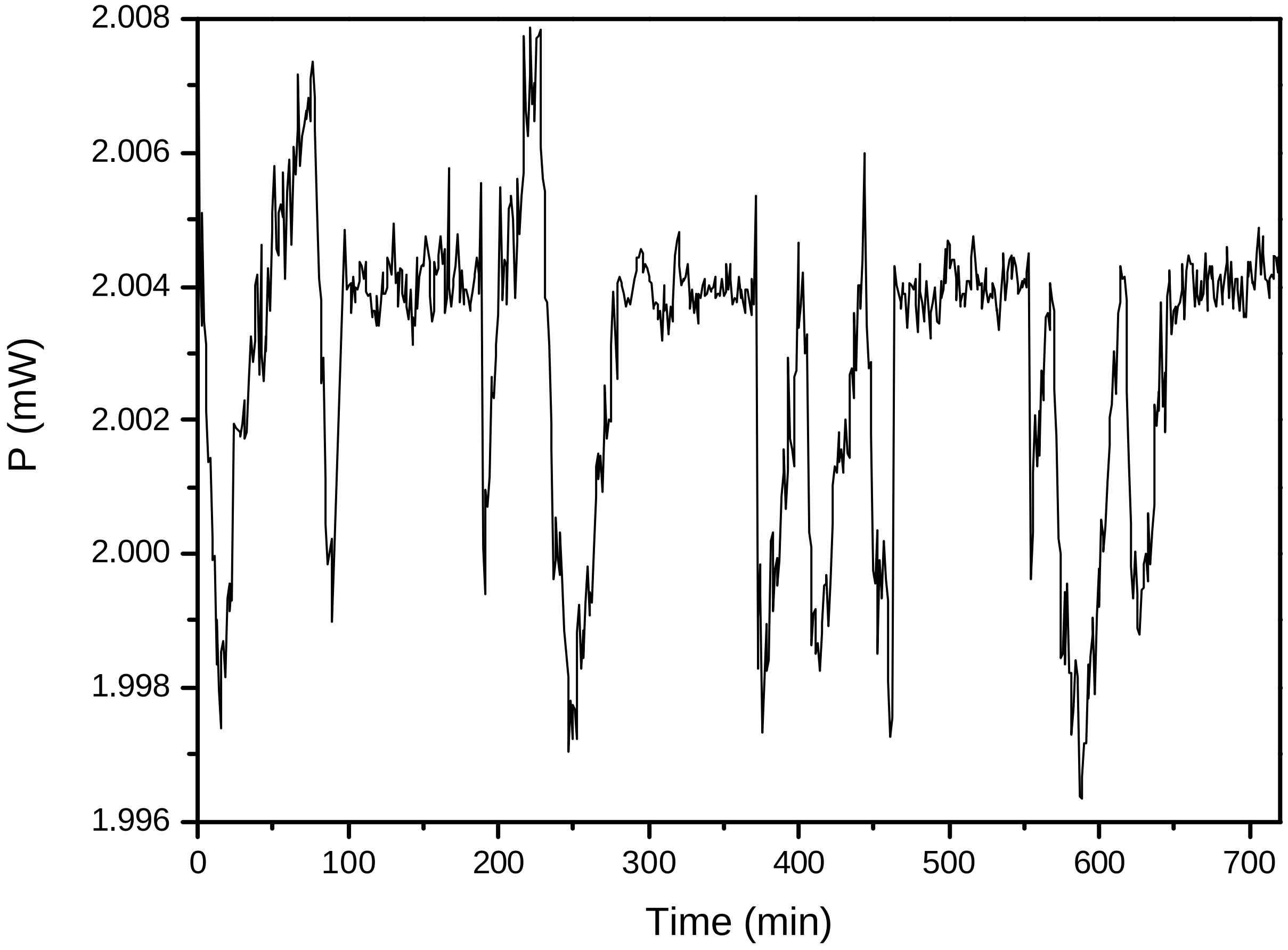

Figure 3

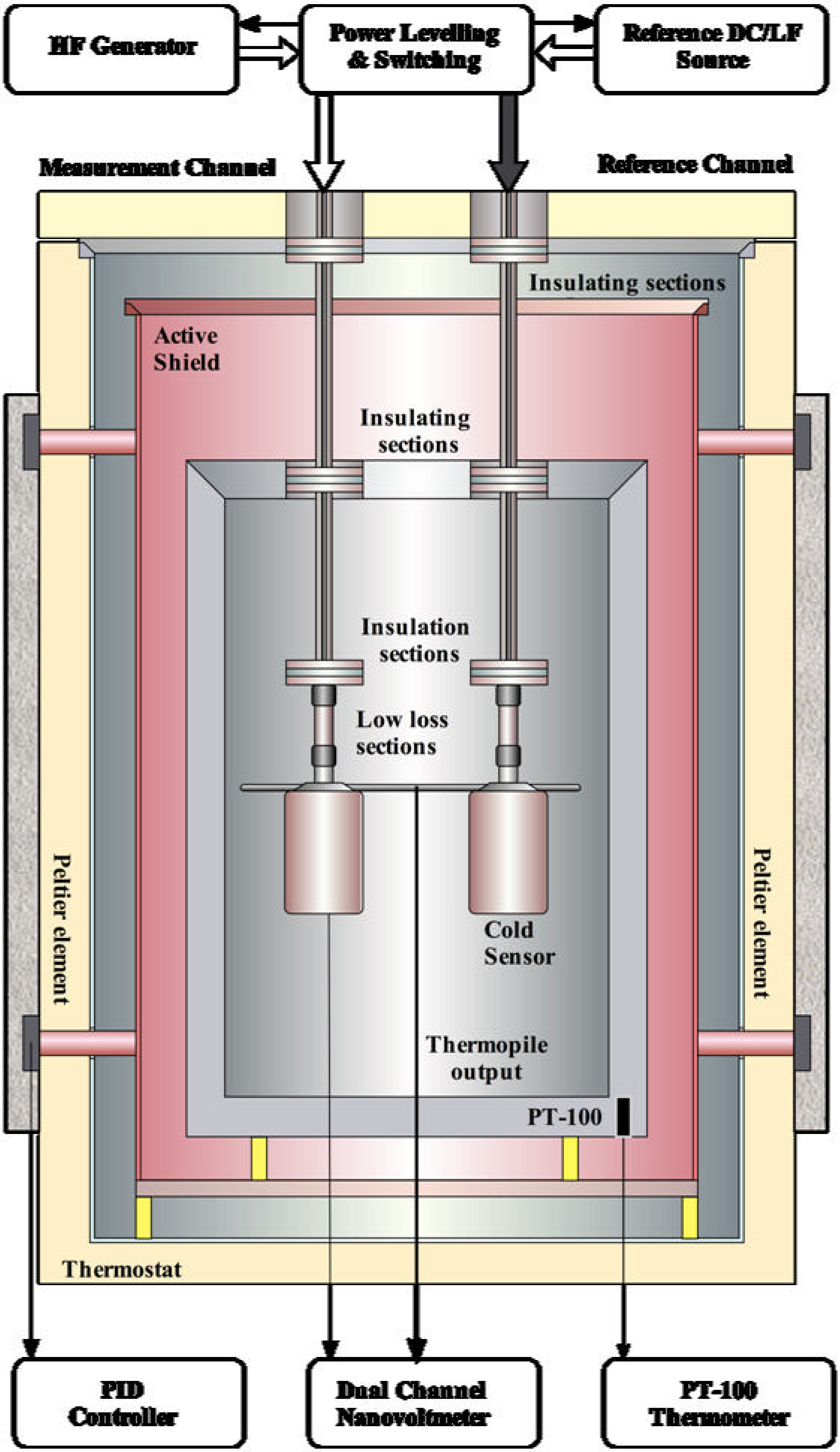

**Figure 4**

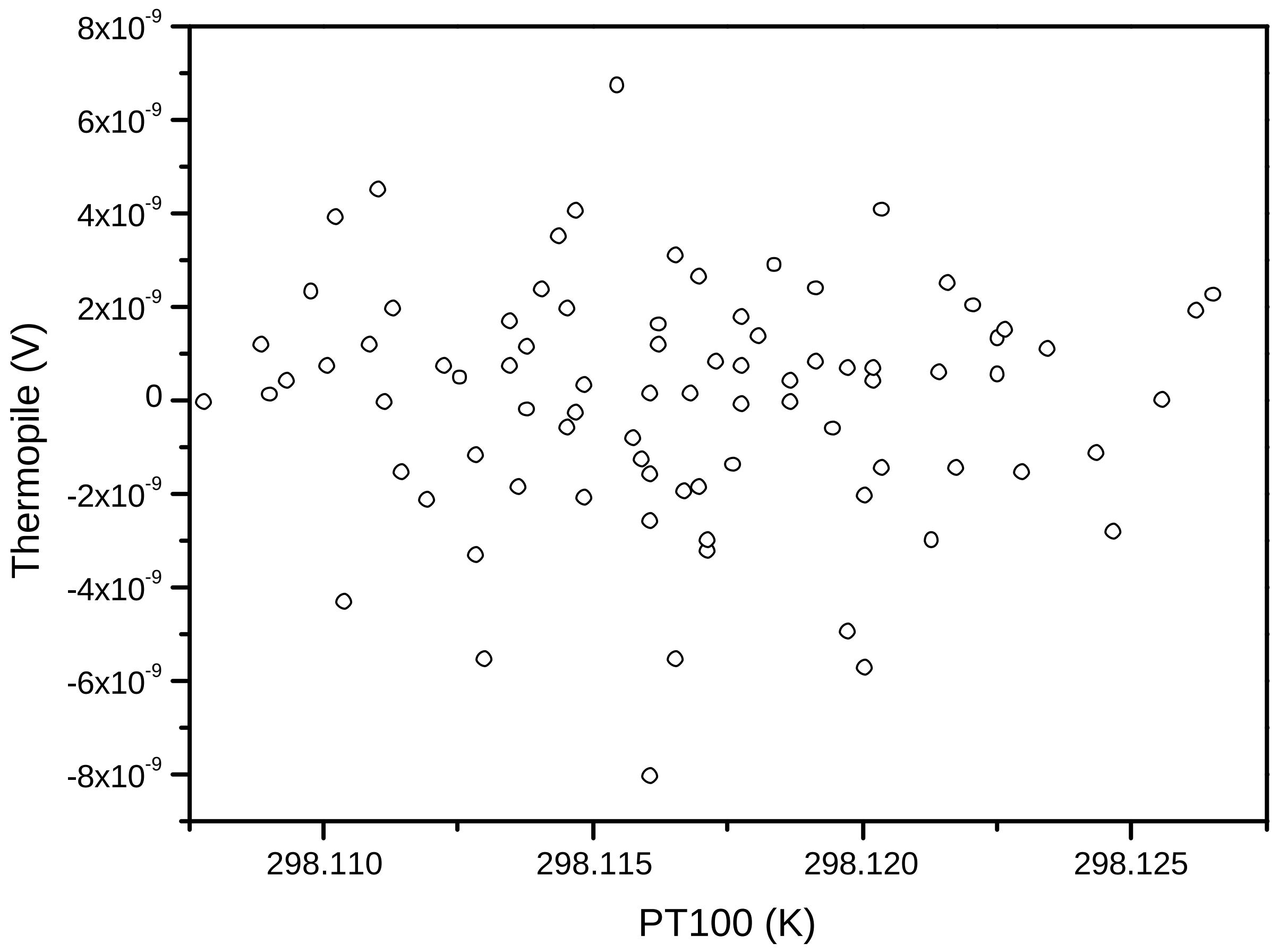

**Figure 5**

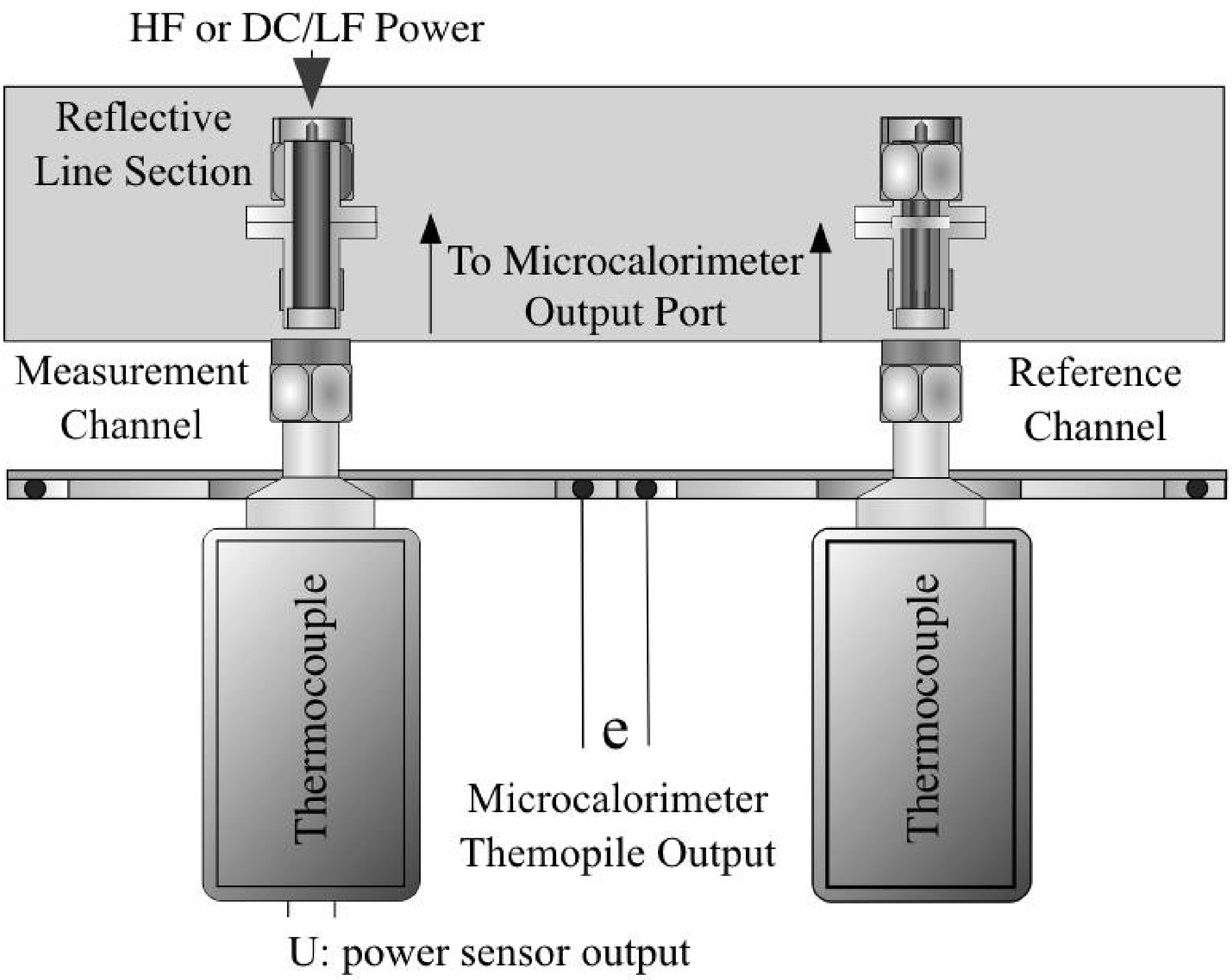

**Figure 6**

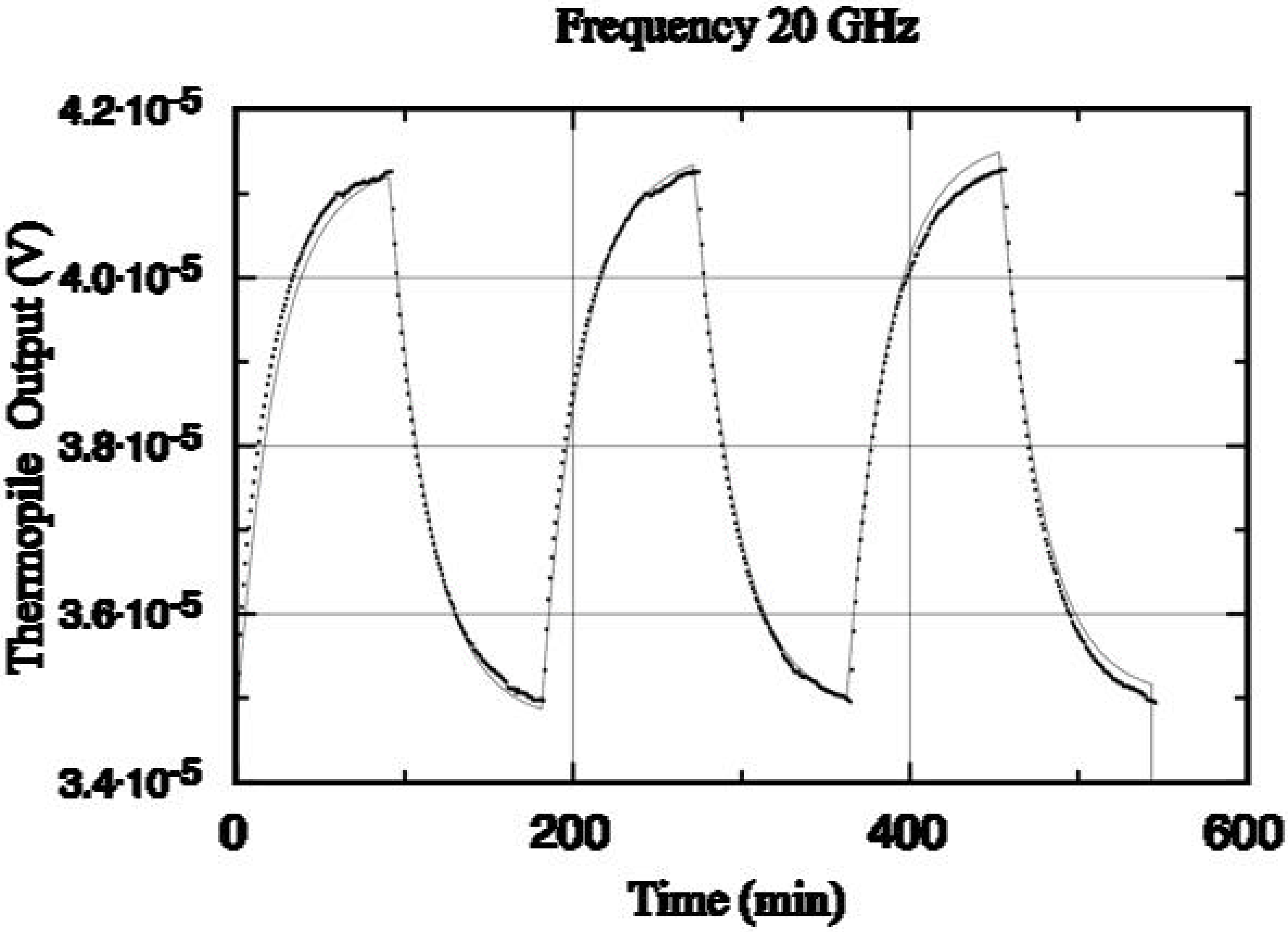

**Figure 7**

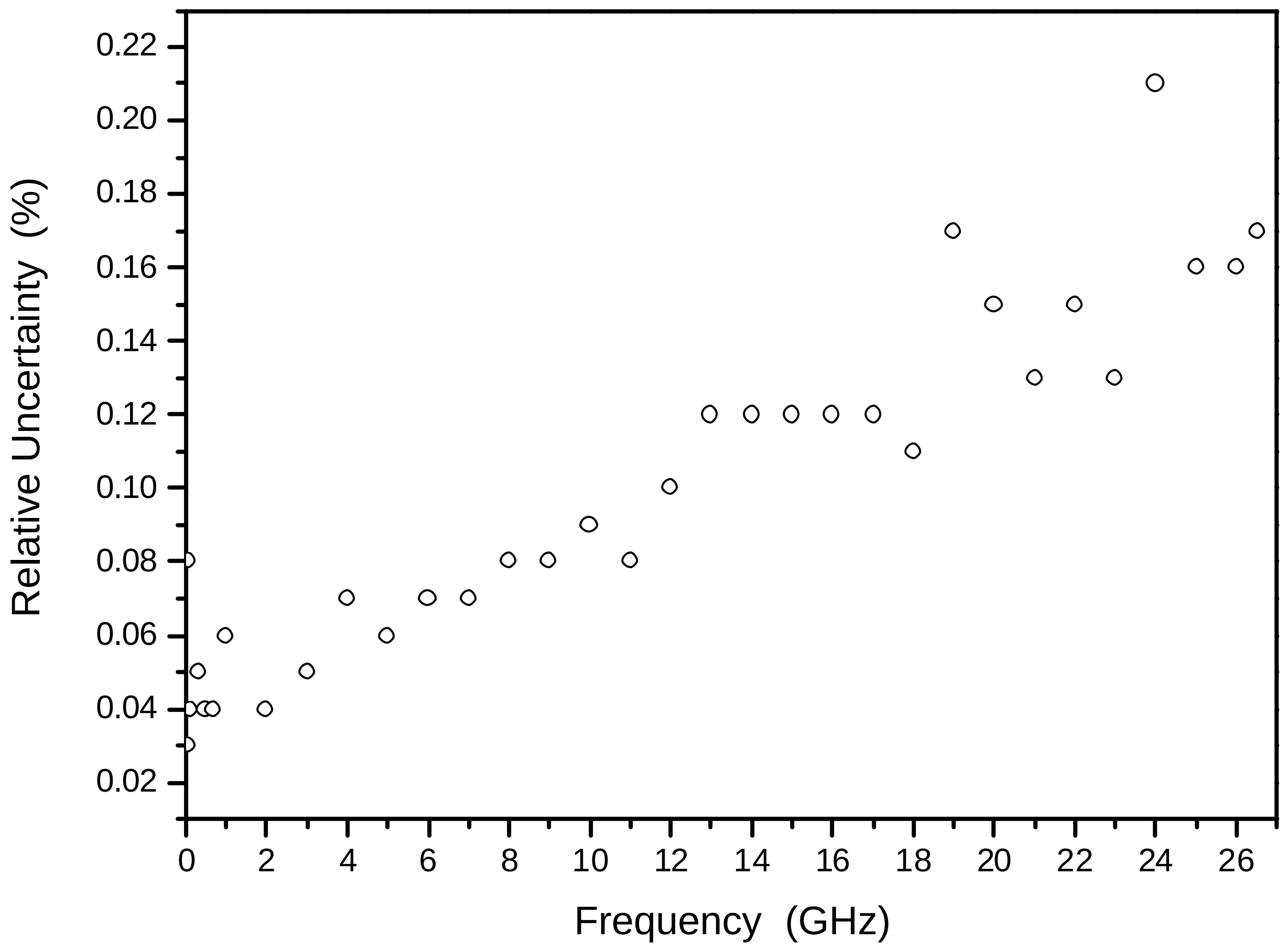